# Tunnel spin injection into graphene using $Al_2O_3$ barrier grown by atomic layer deposition on functionalized graphene surface

Takehiro Yamaguchi[1,a)], Satoru Masubuchi[1,2], Kazuyuki Iguchi[1], Rai Moriya[1,b)], and Tomoki Machida[1,2,3]

[1] *Institute of Industrial Science, University of Tokyo, 4-6-1 Komaba, Meguro-ku, Tokyo 153-8505, Japan*

[2] *Institute for Nano Quantum Information Electronics, University of Tokyo, 4-6-1 Komaba, Meguro-ku, Tokyo 153-8505, Japan*

[3] *PRESTO, Japan Science and Technology Agency, 4-1-8 Honcho, Kawaguchi 332-0012, Japan*

We demonstrate electrical tunnel spin injection from a ferromagnet to graphene through a high-quality $Al_2O_3$ grown by atomic layer deposition (ALD). The graphene surface is functionalized with a self-assembled monolayer of 3,4,9,10-perylene tetracarboxylic acid (PTCA) to promote adhesion and growth of $Al_2O_3$ with a smooth surface. Using this composite tunnel barrier of ALD-$Al_2O_3$ and PTCA, a spin injection signal of ~30 Ω has been observed from non-local magnetoresistance measurements at 45 K, revealing potentially high performance of ALD-$Al_2O_3$/PTCA tunnel barrier for spin injection into graphene.

---

a) E-mail: yamatake@iis.u-tokyo.ac.jp
b) E-mail: moriyar@iis.u-tokyo.ac.jp

1. **Introduction**

   Owing to its unique electrical properties and band structure, graphene has recently received much attention from the perspectives of fundamental and applied physics[1]. Graphene is a particularly promising material for application to spintronics because it is expected to have a very long spin-diffusion length due to its weak spin-orbit and hyperfine interactions[2-5]. In the spintronics, electrical spin injection from a ferromagnetic electrode (FM) to single-/multi-layer graphene is a key technology and it has recently been studied by several research groups[6-9]. Toward more efficient spin injection, various type of interfaces between FM and graphene have been examined, including direct deposition of FM on graphene, reduction of contact area, insertion of a low-temperature-grown $Al_2O_3$ tunnel barrier, and a molecular beam epitaxy (MBE)-grown MgO tunnel barrier[6-10]. Very recently, a large spin transport signal has been observed by non-local magnetoresistance (NLMR) measurements when the tunnel barrier is inserted between the FM and graphene[10,11]. On the other hand, there is a strong demand for further improvement of spin injection efficiency, spin life time, and quality of the FM/graphene interface because the observed spin life time is still shorter than theoretical expectations[3-5] and the spin injection yield is strongly influenced by the fabrication method[12]. Thus, it is necessary to explore more reliable fabrication methods and alternative types of tunnel barrier material for fabricating uniform and pinhole-free tunnel barriers on the graphene surface.

   Atomic layer deposition (ALD) is an emerging new technology for depositing smooth and high-quality oxide films. Given its self-limiting growth process, ALD has the potential to control the thickness of deposited layers on the order of the atomic level.

Thus, ALD is a very promising method to fabricate thin tunnel barriers on the graphene surface. On the other hand, a clean graphene surface is inert for ALD precursors because there are no available dangling bonds. Recently, however, ALD of $Al_2O_3$ on graphene surfaces has been demonstrated using non-covalent surface functionalization layers[13-15]. Among these, the 3,4,9,10-perylene tetracarboxylic acid (PTCA) self-assembled monolayer (SAM)[15] [Fig. 1(a)] is advantageous for the fabrication of tunnel barriers on graphene. In this letter, we demonstrate tunneling spin injection into graphene using an ALD-fabricated $Al_2O_3$ barrier on a graphene surface functionalized with PTCA SAM.

2**. Experiments**

Single- and multi- layer graphene is fabricated by mechanically exfoliating Kish graphite and transferring it to $SiO_2$/n-Si substrate. Subsequently, ALD-$Al_2O_3$/PTCA is fabricated by the following sequence of steps[15]. First, the sample is annealed at 600 °C in vacuum with 1 Torr argon atmosphere for surface cleaning. This produces a clean graphene surface with an RMS roughness of ~0.15 nm. Second, after the sample is dipped in the PTCA solution, some of the physically absorbed PTCA molecules are removed by rinsing with methanol and water. The graphene surface is covered with PTCA SAM at the end of this step, as illustrated in Fig. 1(a). Third, $Al_2O_3$ is fabricated by ALD using trimethyl aluminum (TMA) and water vapor as precursors at sample temperature of 100 °C. The -OH groups of PTCA act as bonding sites for TMA precursor gas. This enables the deposition of TMA, which is subsequently oxidized by water vapor and converted to an Al-O atomic layer. The pulse/purge times used for TMA and water vapor are 0.1 s/15 s and 0.1 s/40 s, respectively. These conditions provide a growth rate of 0.12–0.15 nm/cycle for $Al_2O_3$. The RMS roughness of the surface after 11 cycles of $Al_2O_3$ deposition is ~0.17 nm; this is about the same as the RMS roughness before $Al_2O_3$ deposition.

After the graphene and $SiO_2$ on the substrate surface is completely covered with $Al_2O_3$, ferromagnetic permalloy (Py = $Ni_{81}Fe_{19}$) and non-magnetic Au/Ti electrodes are fabricated with standard electron beam (EB) lithography (Elionix ELS-7500) and EB evaporation. Figures 1(b) and 1(c) show an illustration and photograph, respectively, of the fabricated device structure. Four Py (Py-1~Py-4) and two Au/Ti electrodes are fabricated on the same graphene flake. The length between each Py electrode $L$ is ~1.2 μm. Note that a back-gate voltage $V_{BG}$ can be applied to the n-Si substrate to change the carrier density of the graphene.

## 3. Results and discussion

Prior to the spin injection experiment, we studied the change in contact resistance $R_J$ between FM and graphene in relation to the thickness of $Al_2O_3$. We fabricated many devices with different numbers of $Al_2O_3$ ALD cycles and evaluated $R_J$ based on three terminal measurements at room temperature (RT). The $R_J$ measurement results for which $V_{BG}$ is set to 0 V are plotted in Fig. 1(d). Note that we plotted $R_J$ data obtained from both single- and multi-layer graphene (2–10 layers) because $R_J$ is independent of the number of layers[16]. Although there is a scattering on $R_J$ due to the un-optimized tunnel barrier fabrication process, it can be seen that $R_J$ tend to increases with the cycle number using ALD on PTCA. This suggests that $Al_2O_3$/PTCA barrier is formed between FM and graphene, and its layer thickness could be controlled with cycle number, after more careful optimization. To see the effect of $Al_2O_3$ and PTCA on the properties of graphene, we measured the sheet resistance $R_{sq.}$ of a single-layer graphene on a device with 11 ALD cycles and a contact resistance of ~150 kΩ [Fig. 1(e)] by varying the $V_{BG}$. The width of the graphene channel $W$ is about 1.5 μm. The Dirac point $V_D$ is determined to be +25 V. The mobility of this device, which is determined as 2500 cm$^2$/V, is about the same with or without the $Al_2O_3$/PTCA layer if it is fabricated using the same recipe except for $Al_2O_3$/PTCA deposition. Hence, the composite barrier of $Al_2O_3$/PTCA does not appear to degrade the transport property of the graphene layer.

The circuit for measuring the NLMR is also schematically drawn in Fig. 1(b). Bias is applied between the left-Py and left-Au electrodes to inject spin-polarized electrons from Py into graphene. The spin diffusion in the graphene layer is detected as a generated voltage between the right-Py and right-Au/Ti electrodes. Devices are fabricated with Py electrodes of two different widths; Py-1 and Py-3 are 400 nm wide, and Py-2 and Py-4 are 180 nm wide [Fig 1(c)]. Py electrodes exhibit different coercive forces at different widths, thus allowing an antiparallel magnetic configuration of the electrodes to be realized. NLMR is measured under an external magnetic field $B_{(//, \perp)}$ for various $V_{BG}$, at a measurement temperature of 45 K using a d.c. current of 0.1 μA. The NLMR curve measured under in-plane field $B_{//}$ using the above-mentioned 11-cycle device at $V_{BG}$ = 15 V is shown in Fig. 2(a). A clear NLMR signal is observed, and the change in non-local magnetoresistance $\Delta R_{NL}$ is up to ~30 Ω at 45 K and this value reduces to about half at room temperature. Even though tunnel barrier fabrication process is not fully optimized yet, NLMR signal of exceeding ~10 Ω is obtained. If we take into account the fact that this is the first demonstration of ALD-$Al_2O_3$ as a tunnel spin injector, we think the observed amplitude of NLMR is very promising compared with the previously observed NLMR on a single-layer graphene[10]. The Hanle effect on the same device is also measured using a out-of-plane field $B_\perp$; the obtained data for perpendicular magnetic configurations at $V_{BG} = 0$ V are shown in Fig. 2(b). For the Hanle effect we used lock-in detection with an a.c. current of 0.3 µA to improve the signal-to-noise ratio. As a result, the amplitude of $\Delta R_{NL}$ decreases because $\Delta R_{NL}$ depends on the current amplitude. The data are analyzed based on a method previously used for graphene spin valves[17], and the fitted curve is plotted together with the data in Fig. 2(b). From this analysis, the extracted injected spin polarization, spin relaxation time, spin diffusion constant, and spin diffusion length are $P = 0.06$, $\tau_s = 175$ ps, $D_s = 0.1$ m$^2$/s, and $\lambda_G = 4.18$ µm, respectively. The long spin diffusion length presumably originates from suppression of the interaction between graphene and FM due to the large contact resistance[10], although development of a detailed understanding of the obtained $P$, $D_s$, and $\tau_s$ is still in progress. These NLMR and Hanle effect are strong evidence for spin injection into the graphene through the ALD-Al$_2$O$_3$/PTCA composite barrier and spin transport across the graphene channel.

We summarized the relation between $\Delta R_{NL}$ and $R_J$ at $V_{BG} = 0$ V obtained from single-layer, bi-layer, and multilayer graphene in Fig. 2(c). The $\Delta R_{NL}$ increases with $R_J$ for single-layer graphene, suggesting that control of contact resistance $R_J$ is crucial to obtain a large $\Delta R_{NL}$. The ALD technique enables control of $R_J$ over a wide range of values simply by adjusting the cycle number. To confirm whether we achieved tunnel spin injection into graphene, we examined the dependence of $\Delta R_{NL}$ on $V_{BG}$ for the device which has highest $\Delta R_{NL}$. According to a simple spin drift-diffusion model, the shape of the function relating $\Delta R_{NL}$ and $V_{BG}$ dramatically changes with the balance between the spin resistance of graphene $R_G$ and the contact resistance $R_J$[10,18]. From ref. 18, $\Delta R_{NL}$ is formulated as follows;

$$\Delta R_{NL} = 4R_G e^{-\frac{L}{\lambda_G}} \left( \frac{P_J \frac{R_J}{R_G}}{1-P_J^2} + \frac{P_F \frac{R_F}{R_G}}{1-P_F^2} \right)^2 \times \left( \left(1 + \frac{2\frac{R_J}{R_G}}{1-P_J^2} + \frac{2\frac{R_F}{R_G}}{1-P_F^2}\right)^2 - e^{-\frac{2L}{\lambda_G}} \right)^{-1}, \quad (1)$$

where $R_G = R_{sq.}\lambda_G/W$ and $R_F = \rho_F\lambda_F/A_J$ are the spin resistance of graphene and ferromagnet, respectively, $R_{sq.}$ is the sheet resistance of graphene, $\rho_F$ is the resistivity of ferromagnet, $\lambda_G$ and $\lambda_F$ are the spin diffusion length of graphene and ferromagnet, respectively, $A_J$ is the contact area of the junction, $P_J$ is the injected spin polarization, $P_F$ is the spin polarization of ferromagnet, respectively. For the typical graphene devices, $R_F \ll R_G, R_J$. Under this condition, $R_J$ is classified into three different regions: $R_J \gg R_G$ (tunneling), $R_J \approx R_G$ (intermediate), and $R_J \ll R_G$ (transparent)[10]. We plot representative curves for these regions in Fig. 3(b) using device parameters obtained from our device[19]. When $R_J$ increases, the function relating $\Delta R_{NL}$ and $V_{BG}$ shows a clear transition from a valley shape (transparent) to a peak shape (tunneling). The data displayed in Fig. 3(a) show good agreement with the case of tunneling contact [see top panel of Fig. 3(b)]. In addition, the $R_J$ used in the calculation (~100 kΩ) is very close to the $R_J$ of the device shown in Fig. 3(a). The *I-V* characteristic of the contact between Py and graphene is evaluated based on three terminal measurements and is shown in the inset of the Fig. 3 (a). The non-linear tunnel barrier-type *I-V* curve is observed. These indicates that a high-quality tunnel barrier is formed with the ALD-grown $Al_2O_3$ barrier. The discrepancy between the measurement data and equation (1) could be due in part to changes in the spin diffusion length with $V_{BG}$.

## 4. Summary

We demonstrated electrical spin injection from a ferromagnet to graphene using an ALD-$Al_2O_3$/PTCA composite tunnel barrier. With the help of PTCA SAM, the adhesion of $Al_2O_3$ during the ALD process is greatly improved. The results showed that tunnel spin injection is achieved and that a large non-local magnetoresistance signal is obtained. These performance characteristics of the $Al_2O_3$/PTCA composite tunnel barrier have potential applications for future graphene spintronics.

## 5. Acknowledgement

We are grateful for helpful discussions with K. Yamaguchi. We also thank K. Hirakawa and K. Shibata for access to ALD. This work is supported by the PRESTO, Japan Science and Technology Agency, the Special Coordination Fund for Promoting Science and Technology, the Grant-in-Aid for Scientific Research from the MEXT, and the CREST, Japan Science and Technology Agency.

**Figure Captions**

Figure 1

(a) Schematic illustration of atomic layer deposition (ALD) of $Al_2O_3$ on a clean graphene surface functionalized with PTCA self-assembled monolayer. (b) Illustration of the structure and non-local magnetoresistance measurement circuit for a graphene spin valve device. (c) Photograph of the fabricated device. Graphene flakes are indicated by the dotted line. Six metal electrodes are fabricated on graphene; these are non-magnetic Au/Ti and ferromagnetic Py (Py-1~Py-4) electrodes. (d) Relations between the $R_J$ and ALD cycle number are plotted as solid circles at RT and $V_{BG} = 0$ V. Data are taken from the single-/multi-layer graphene with layer numbers between 1–10. The red line is a visual guide. (e) Dependence of the graphene sheet resistance $R_{sq.}$ on the back-gate voltage $V_{BG}$.

Figure 2

(a) Non-local magnetoresistance (NLMR) of the graphene spin valve with 11 cycles of ALD-$Al_2O_3$ for $V_{BG} = 15$ V ($V_D = 25$ V). (b) Black and gray dots indicate the Hanle curve of the same device at parallel (P) and anti-parallel (AP) magnetic configuration, respectively. Solid green and blue line shows a fitted curve based on ref. 17. (c) The relation between $\Delta R_{NL}$ at 45 K and the geometric mean of contract resistances for the injector $R_{J\ inj}$ and the detector $R_{J\ det}$.

Figure 3

(a) Changes in the $\Delta R_{NL}$ signal with the back-gate voltage measured at 45 K. Insets: *I-V* characteristic of the contact between injector Py and graphene at the $V_{BG}$ = 0 V. (b) Simulations of the function relating $\Delta R_{NL}$ to $V_{BG}$ based on eq. (1) with measured $R_{sq.}$ values and given $R_J$ values. From the top to bottom panels, the $R_J$ is 100, and 1, and 0.01 kΩ, respectively.

Figure 1

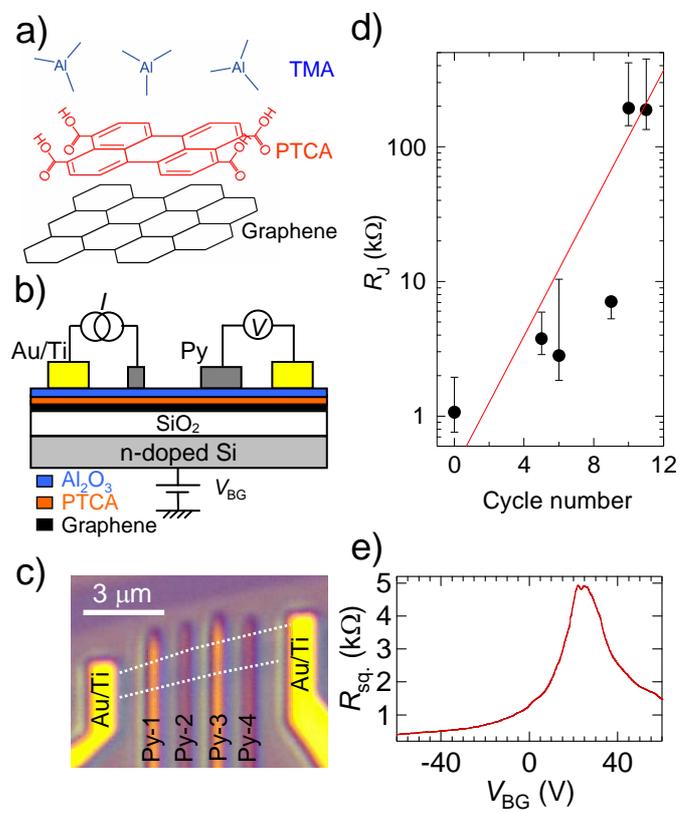

Figure 2

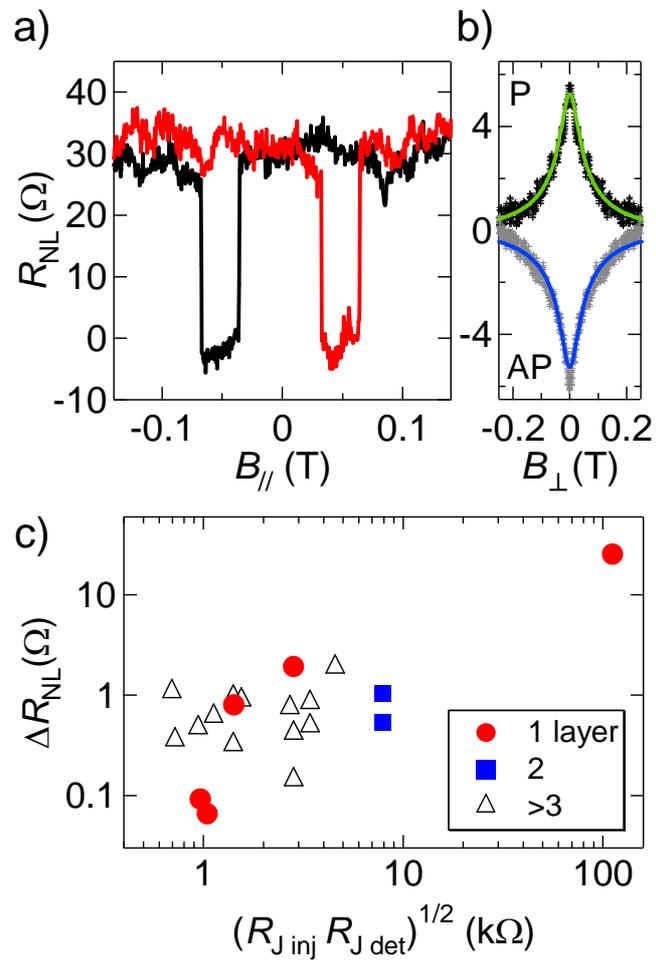

Figure 3

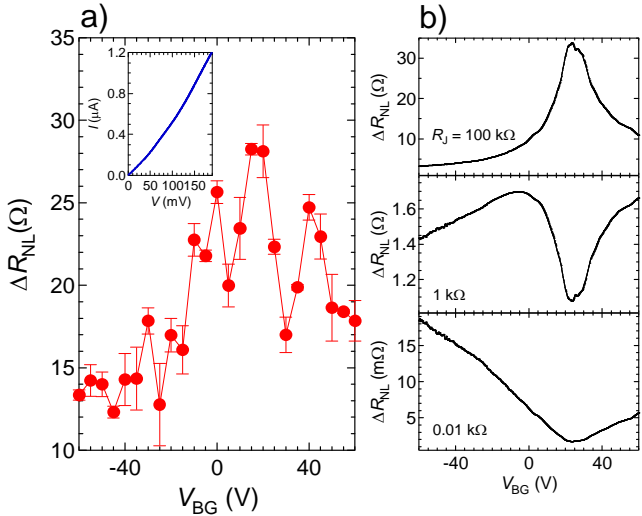